\documentclass[prl, twocolumn]{revtex4}
\usepackage{graphicx}
\usepackage{hyperref}
\usepackage{ulem}
\usepackage[section]{placeins}
\usepackage{amsmath}
\usepackage{amssymb}
\usepackage{dsfont}
\usepackage{bm}
\usepackage{color}

\usepackage{mathtools}
\DeclarePairedDelimiter{\ceil}{\lceil}{\rceil}

\usepackage{algorithm}
\usepackage[noend]{algpseudocode}

\makeatletter
\def\BState{\State\hskip-\ALG@thistlm}
\makeatother

\newcommand{\ket}[1]{|#1\rangle}             % Ket Dirac's notation %
\newcommand{\bra}[1]{\langle#1|}             % Bra Dirac's notation  %
\begin{document}

\title{Projected gradient descent becomes general}
\title{{Projected gradient descent algorithms for quantum state tomography}}
\author{ Eliot~Bolduc$^{1}$, George Knee$^2$, Erik Gauger$^1$, Jonathan~Leach$^{1}$}
\email[email: ]{J.Leach@hw.ac.uk}
\affiliation{$^1$ SUPA, Institute of Photonics and Quantum Sciences, Heriot-Watt University, David Brewster Building, Edinburgh, EH14 4AS, UK}
\address{$^2$ Department of Physics, University of Warwick, Coventry, CV4 7AL, UK}

\date{\today}

\begin{abstract}
Accurate quantum tomography is a vital tool in both fundamental and applied quantum science. 
It is a task that involves processing a noisy measurement record in order to construct a reliable estimate of an unknown quantum state, and is central to quantum computing, metrology, and communication. To date, many different approaches to quantum state estimation have been developed, yet no one method fits all applications, and all fail relatively quickly as the dimensionality of the unknown state grows.  In this work, we suggest that projected gradient descent is a method that can evade some of these shortcomings. We present three novel tomography algorithms that use projected gradient descent and compare their performance with state-of-the-art alternatives, i.e.~the diluted iterative algorithm and convex programming. Our results find in favour of the general class of projected gradient descent methods due to their speed, applicability to large states, and the range of conditions in which they perform.

\end{abstract}
\maketitle

\section{Introduction}

The reconstruction of an unknown quantum state {-- known as quantum tomography --} is a fundamental task in quantum information science, where a myriad of new technologies promise to exploit special features of quantum states in order to enhance communication, metrology, and computation. Since the quantum state represents maximal information about a physical system, all physical properties can be calculated from it.  Checking for the existence for a highly entangled state, a state which can violate a Bell inequality, or even the initial state required in a gate-based quantum computer are thus just some examples of the importance of inferring the quantum state from laboratory data. As experimental methods progress, the complexity of quantum systems that can be well controlled in the laboratory grows. {In recent times, for example, various groups have been able to demonstrate quantum control of a high number of qubits ~\cite{Gao:2010ww,Schindler:2011,Yao:2012uy}.} To gain an idea of the challenge of state reconstruction, one need only consider that the number of real parameters required to describe the joint state of $n$ qubits scales as order $2^{2n}$. Alternatively, the orbital angular momentum of single photons, for example, is a single degree of freedom with a large amount of internal structure: it has recently been characterised via the reconstruction of a 100,000 dimensional statevector~\cite{Bolduc:2016, Malik:2013, Wong:2012}.  Quite apart from the challenges presented by preparation and measurement of quantum states, tackling the state reconstruction problem in the face of such complexity calls for sophisticated data processing techniques, {which are the focus of this paper.}

{ \paragraph{} Tomography experiments involve subjecting a system, described by some unknown quantum state, to a well-defined measurement procedure and recording the measurement outcome. The central tenets of quantum theory place severe restrictions on one's ability to characterise an unknown quantum system given measurements made on only a single copy. One assumes, therefore, access to a large but finite number of copies of a system, all prepared in an identical quantum state. As the complexity of the quantum state grows, the number of detector counts for each state parameter necessarily shrinks. Particularly in optical systems, the prevalence of low detection efficiency exacerbates this problem, leaving the tomographer with a noisy data-set from which to make her inferences. In an idealised situation where all noise (including statistical) is absent, the true state $\rho_\text{true}$ can be found exactly. Here we concentrate on the more realistic situation, and assume only that the measurement procedure is informationally complete (that is to say, the measurement record contains a nonzero amount of information available about each quantum state parameter), and turn our attention to the question of processing this data toward the most likely estimate of the unknown state.}

 \paragraph{} In most cases, quantum tomographers must employ numerical techniques to search for the best estimate possible, given the data that has been collected. In this work, we analyse the algorithmic method of projected gradient descent as applied to quantum tomography, and benchmark it against a number of existing methods. The state-of-the-art with regards to full quantum tomography methods include the diluted iterative algorithm (DIA)  \cite{Rehacek:2001,Rehacek:2007} and convex programming \cite{Grant:2008}. Both methods benefit from a theoretical guarantee: that they converge to the maximum likelihood (ML) state $\rho_\text{ML}$ (discussed below). However, the DIA has been observed to converge slowly \cite{Goncalves:2016, Silva:2016}, and convex programming solvers such as SDPT3 and SeDuMi are known to require computational time that scales poorly with non-sparse matrix dimensionality \cite{Toh:1999, Sturm:1999}.

 \paragraph{}     {A non-iterative quantum tomography method was devised by Smolin \textit{et al.}, who showed that, if the measurement operators are traceless and the noise is of the Gaussian type, the constrained maximum likelihood state $\rho_\text{ML}$ can be retrieved in a single projection step from the unconstrained maximum likelihood state $\chi_{ML}$, which can be found very quickly using linear inversion \cite{Smolin:2012vp}. With an implementation of this method's core algorithm using a GPU, Guo \textit{et al.}~were able to recover a simulated 14-qubit density matrix \cite{Guo:2016tz}.} {However, the restrictive above conditions motivate the search for more broadly applicable efficient techniques.} 
   
{ Projected gradient descent (PGD) methods are emerging as promising candidates for quantum tomography  \cite{ Goncalves:2016, Shang:2016tva}. We present three PGD algorithms that converge towards the maximum likelihood quantum state}: projected gradient descent with backtracking (PGDB) \cite{Goncalves:2016, Shang:2016tva}, Fast Iterative Shrinkage-Thresholding Algorithm (FISTA) \cite{Beck:2009tk} and projected gradient descent with momentum (PGDM). We provide evidence that they converge faster than both DIA and SDPT3, and scale more favourably than SDPT3.  {We also find that the PGD methods are very versatile in that one can model a wide variety of types of noise; in particular, the case of ill-conditioned measurements.}

{Gon\c calves \textit{et al.}~\cite{ Goncalves:2016} and Shang \textit{et al.}~\cite{Shang:2016tva} have both recently discussed PGDB as an efficient technique for quantum tomography: however, the algorithmic variants we introduce here can significantly outperform it. Furthermore, Shang \textit{et al.} considered Pauli measurements, for which the technique from Ref.~\cite{Smolin:2012vp} is highly efficient. It is therefore vital to study the performance of projected gradient descent outside of this realm to establish its true usefulness. We here confirm that PGD methods continue to exhibit excellent properties in scenarios where the technique from Ref.~\cite{Smolin:2012vp} is not applicable. }

 \begin{figure}[t!]
  \centering
  \includegraphics[width=0.49 \textwidth]{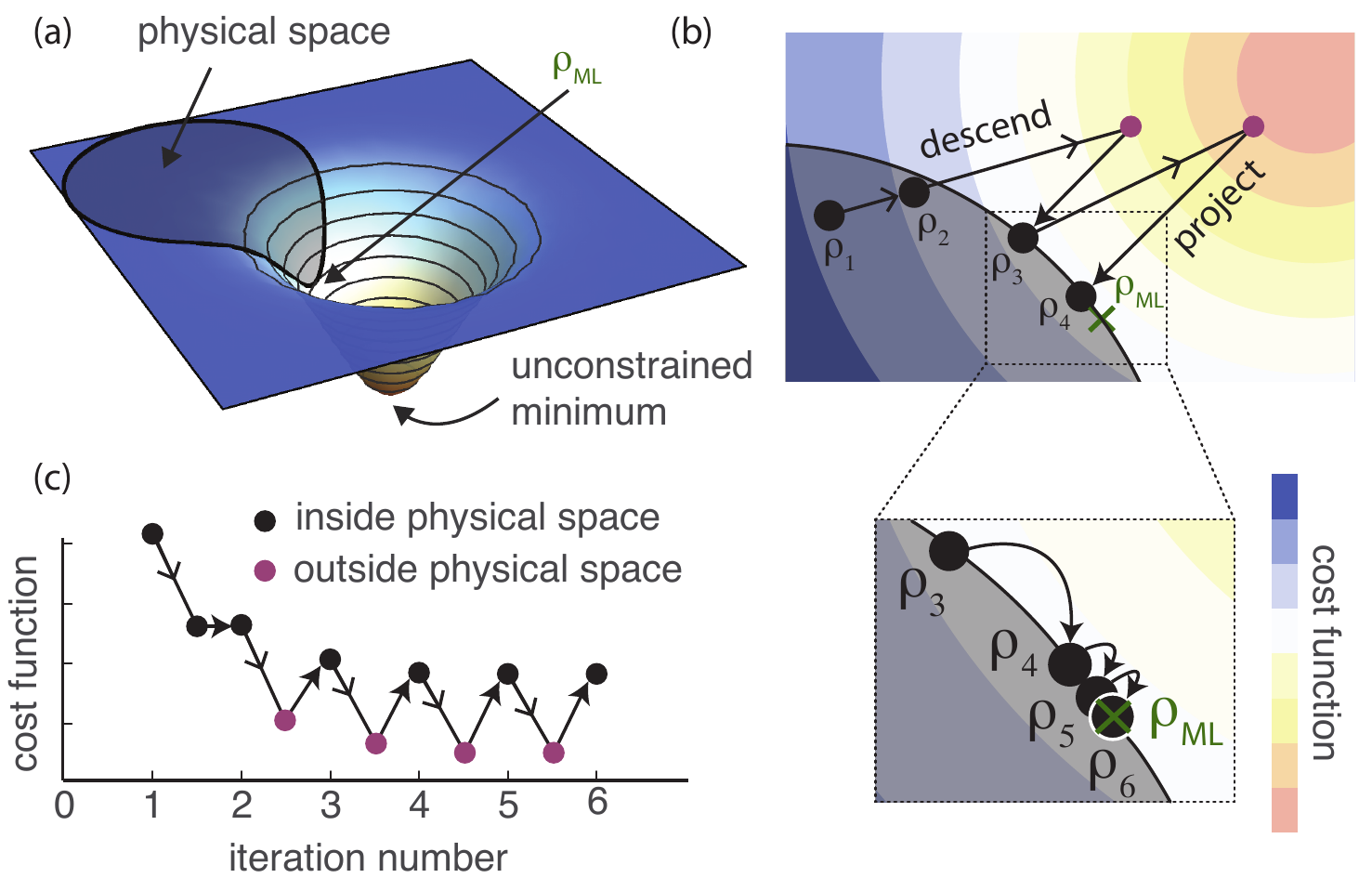} 
  \caption{\footnotesize{a) Schematic showing the physical space as a convex subset of the set of unconstrained  matrices, The minimum of the cost function is often outside of the physical space. b) Illustration of the PGD process for a qubit. A step in the gradient direction can yield a non physically-allowed density matrix, but the projection brings the estimate back into the desired search space, (i.e. the Bloch sphere in the case of a qubit). c) This process lowers the cost function, and is repeated until progress is sufficiently small and the final density matrix estimate is as close as desired to the maximum likelihood state $\rho_\text{ML}$. }}
  \label{fig:scheme}
\end{figure}

  \paragraph{} The remainder of this paper is structured as follows: After giving an introduction to quantum tomography and the general idea of projected gradient descent, we lay out three PGD algorithms and discuss their performance: namely, convergence profiles and running time. {To the best of our knowledge, FISTA has never previously been applied to quantum tomography and PGDM is a novel algorithm altogether}.  DIA and SDPT3, considered as current state-of-the-art algorithms, will serve as benchmarks by which the PGD approach will be judged. Finally, we report on the results of state reconstruction using pseudo-experimental data, i.e. simulations of realistic quantum tomographic experiments with noise. We consider three figures of merit for quantum tomographic techniques: the convergence time and the fidelity between the recovered state and the actual one $\rho_\text{true}$. Over a broad range of Hilbert space dimensions, we run the algorithms multiple times, each time with a randomly generated density matrices with fixed purity \cite{Zyczkowski:2011}, and record the running times and fidelities.
  
{ \subsection{Quantum tomography}}
The Born rule,
%\begin{align}
$ p_i = \text{Tr}(\Pi_i\rho)$,
%\end{align}
 being the central equation of quantum theory, encodes the probability $p_i$ to obtain a certain measurement outcome given a particular quantum state. It involves a quantum state $\rho$, which takes the form of a $d\times d$ positive-semidefinite matrix of unit trace, and an Hermitian measurement operator $\Pi_i$.  
 Quantum tomography is essentially the process of finding the density matrix whose calculated outcome probabilities {(for an informationally complete set of $N$ operators)} most closely match the experimentally observed data. 
 
\paragraph{} The probabilities $p_i$ are of course not directly observable, only the number of clicks $n_i$ recorded in a real measurement device after a finite number of trials. In the absence of noise, the probabilities relate to the number of clicks through a multiplicative factor $r$: $n_i = r p_i$. In real situations, there is a discrepancy between $r p_i$ and $n_i$ due to i) technical noise in the measurement device and ii) statistical uncertainty. Furthermore, if the noise is particularly severe, the matrix reconstructed with naive methods (such as linear inversion) will fail to qualify as a physical quantum state: the positivity or unit-trace properties can be violated. Multiple techniques have therefore been developed that allow one to search for an estimate matrix that is guaranteed to be physical. Examples include searching for the Cholesky factor $T$ (where $\rho=TT^\dagger$ is guaranteed positive) and using a Lagrange multiplier (to ensure unit trace)~\cite{Banaszek:1999}. However, searching in the factored space can often lead to an ill-conditioned problem and slow convergence  \cite{Shang:2016tva}. As we evidence,  there are advantages to be had by instead allowing the search to temporarily wander into unphysical territory.

\paragraph{} {The measurement operators, the expectation values $p_i$ and the detector clicks $n_i$ can be stacked into a matrix and two vectors, respectively: 
\begin{equation}
\begin{aligned} 
A =  \left( \begin{array}{ccc} \text{vec}(\widehat \Pi_1)^T \\ \vdots \\ \text{vec}(\widehat \Pi_N )^T  \end{array} \right),
\hspace{5pt} \bold{p} =  \left( \begin{array}{ccc} p_1 \\ \vdots \\ p_{N} \end{array} \right), \hspace{5pt} 
\text{and} \hspace{5pt}  \bold{n} =  \left( \begin{array}{ccc} n_1 \\ \vdots \\ n_{N} \end{array} \right), 
\end{aligned}
\end{equation}
where $N$ is the total number of projectors.  With the above notation, the expectation value vector reads  $
\bold{p} = A \hspace{2pt} \text{vec}(\rho)$. The computation of this vector takes $O(N d^2)$ floating-point operations in general, but a lower computational complexity can be achieved when the operators originate from tensor products \cite{Shang:2016tva}. The accuracy of the maximum likelihood state depends significantly on the condition number (which is the ratio of maximum singular value to minimum singular value) of the measurement matrix $A$ \cite{Miranowicz:2014vr}. High condition numbers, which correspond to ill-conditioned measurement matrices, arise in the fields of detector tomography \cite{Feito:2009uy,Lundeen:2009vj} and superconducting artificial atoms \cite{Bianchetti:2010tq}.  }

{The multiplicative factor $r$ can readily be estimated if at least a subset $\mathcal{Z}$ of the measurement matrix $A$ forms a POVM, in which case  the sum of the probabilities belonging to the set $\mathcal{Z}$, $\sum p_\mathcal Z$, is independent of the state $\rho$. For example, if the measurement operators in $\mathcal{Z}$ form a basis, the sum in question amounts to unity. For an arbitrary POVM, the best estimate for the multiplication factor is given by  
\begin{equation}\label{pPOVM}
 r = \frac{d }{N_\mathcal Z}\sum\limits_{j \in \mathcal Z} {n_j} ~,
\end{equation}
  where $N_\mathcal Z$ is the number of projectors in the POVM. Moreover, the average number of clicks per outcome is  $r/d$, and the total number of clicks for this POVM is $r N_\mathcal Z /d$.}

\normalsize
\subsubsection{Summary of PGD algorithms}

\paragraph{} The distance between $p_i$ and $n_i/r$ is to be considered as a `cost function' $\mathcal{C}(\rho)$ in the sense of numerical optimisation. In the minimisation of a cost function, PGD algorithms are useful when one seeks a solution in a proper subset of a larger search space. Since physical quantum states, represented as unit-trace positive-semidefinite matrices, exist in a (convex) subset of the (convex) set of $d\times d$ matrices, quantum tomography is indeed a problem of this kind. Projected gradient descent is an iterative procedure with two substeps. Starting with a well-chosen physical state, first a step is taken in the downhill direction of the cost function, which has the chance to result in a nonphysical matrix. Second, to bring the estimate back within the constrained, physical space, we project it to the closest point in the solution space (for example, using a matrix norm). This two step process is then repeated until the cost function converges towards a low enough value. Since we are searching over a convex set, as long as the cost function is a strictly convex function of $\rho$, there will be a unique solution that minimises it. Fig.~\ref{fig:scheme} shows the evolution of the density matrix estimate of a qubit through six iterations of PGD.  

\subsubsection{Maximum likelihood}

We have yet to specifically define the figure of merit for \textit{closeness} between the estimated probabilities and the outcome frequencies. Maximum likelihood analysis provides a principled way to derive such a figure of merit.  When doing statistical estimation, it is necessary to operate within a statistical model or belief system: one approach is Bayesian estimation, which works with iteratively updating such beliefs using Bayes' rule. Here, however, the beliefs are encoded in the likelihood function for a multinomial experiment:
 \begin{align}
 \mathcal{L}(\rho)\propto \prod_i p_i^{n_i} ~.
 \end{align}
Maximizing this function -- i.e.~finding the quantum state $\rho$ that makes the observed data the most likely -- is the most widely applied philosophy for tomography \cite{Banaszek:1999,Kaznady:2009wf,James:2001ut,Rehacek:2001}. Since the maximum-likelihood state $\rho_{\textrm{ML}}=\max_{\rho}\mathcal{L}(\rho)$ is also  the minimum of $-\log \mathcal L$, we can proceed by minimising the second function, and we may ignore any scale or shift by a constant that is independent of $\rho$. We therefore define the cost function 
 \begin{equation}
 \mathcal C(\rho) = - \log \mathcal L(\rho)
 \end{equation}
that we seek to minimise. When the number of trials is large, {this is well approximated by the Poisson-approximated Gaussian likelihood function  $\mathcal C(\rho)  \approx -\log\mathcal{L}_P(\rho)=\boldsymbol{\nu}^T  \boldsymbol{\nu}$ with $\nu_i = (rp_i-n_i)/{\sqrt n_i}$. Assuming Poisson-distributed data, the variance for outcome $i$ is the number of clicks $n_i$. Hence $\nu_i$ corresponds to the ratio of the error to the expected error on outcome $i$. The true density matrix gives an expected negative log-likelihood per outcome $\mathcal C/N$ of unity because of the noise on the outcomes $n_i$}. A value greater than unity indicates a poor density matrix estimate or an incomplete noise model, whereas a value smaller than unity is a sign of overfitting to noise. In general, the maximum likelihood density matrix overfits the data slightly \cite{Press:1996wea}, but one cannot achieve a better estimate in the absence of prior knowledge.

We are now ready to detail the algorithms for reconstructing  the density matrix. In all of the following algorithms, the completely mixed state $\rho_0=I/d$ will serve as the starting point. Our selection of four iterative algorithms are then defined by  a recursion relation relating the density matrix at the next iteration to the matrix at the current iteration. 

\section{RESULTS}
\subsection{Projected Gradient Descent Algorithms}
The process of any PGD algorithm involves steps in the general gradient direction, interspersed with leaps back into the constrained set \cite{Goncalves:2016,Shang:2016tva}. The simplest such algorithm can be written in a single line \cite{Bolduc:2016}:
\begin{equation}
\rho_{k+1} = \mathcal P[\rho_k -\delta \nabla  \mathcal C(\rho_k) ], \label{baba}
\end{equation}
where  $\delta$ is a step size and $\mathcal P[\cdot]$ is a projection onto the set of unit-trace positive matrices, seeking the `closest' unit-trace positive matrix to its argument. Various approaches can be used to to establish what `closest' means {(including operator norms, see Supplementary Information). We adopt the {\it simplex projection} $\mathcal{P}[\cdot]\rightarrow\mathcal S[\cdot]$, which essentially consists of transforming the eigenvalues of the density matrix such that the sum is unity trace and they are all positive \cite{Michelot:1986vk}.   If the multiplicative factor $r$ is known or computed in advance, the version described in detail in Ref.~\cite{Goncalves:2016} applies, otherwise the projection must instead be performed over the space of positive matrices, preserving the trace of the argument \cite{Smolin:2012vp}. }

\paragraph{} We now proceed to discuss three PGD algorithms which are all are extensions of Eq.~(\ref{baba}). 

\subsubsection{Projected Gradient Descent with Momentum}
Here we augment the basic PGD algorithm of Eq.~(\ref{baba}) with a technique borrowed the momentum-aided gradient descent method from the field of machine learning~\cite{Sutskever:2013}. This technique stores a running weighted-average $M_k$ of the log-likelihood gradient. This running average provides a memory of previous descent directions which is used to better estimate the next descent step. The core of this algorithm reads
 \begin{equation}
\begin{aligned}
{M_{k+1}} &=   \zeta_k M_{k} - \gamma_k \nabla  \mathcal C(\rho_k) ~, \\
\rho_{k+1} &= \mathcal S (\rho_k + M_k) ~,
\end{aligned}
\end{equation} 
where $\zeta_k$ codes for the level of `inertia' for the descent direction, and $\gamma_k$ is the step size. In general, these metaparameters depend on the iteration number $k$, but can also be set as constants throughout the algorithm. We provide full pseudo-code for this and the other algorithms in the Methods section, as well as Matlab implementations.

\subsubsection{Fast Iterative Shrinkage-Thresholding Algorithm}

FISTA was first developed in the context of image denoising \cite{Beck:2009tk}, but here we introduce the method for use in quantum state tomography with adapted refinements. In this implementation of PGD, the change in the iterate $\rho_k$ is not always in the descent direction, i.e. the log-likelihood function can go up, but as we shall see it descends much faster on average than the basic gradient descent algorithm from Eq.~(\ref{baba}).  The core of the algorithm is given by 
\begin{equation}
 \rho_{k+1} = \mathcal S\left[\rho_k + \frac{k-2}{k+1}(\rho_k-\rho_{k-1})  - \delta\nabla  \mathcal C(\rho_k) \right] ~,
 \end{equation}
where $\delta$ is a step size. 

\begin{figure*}
  \centering
  \includegraphics{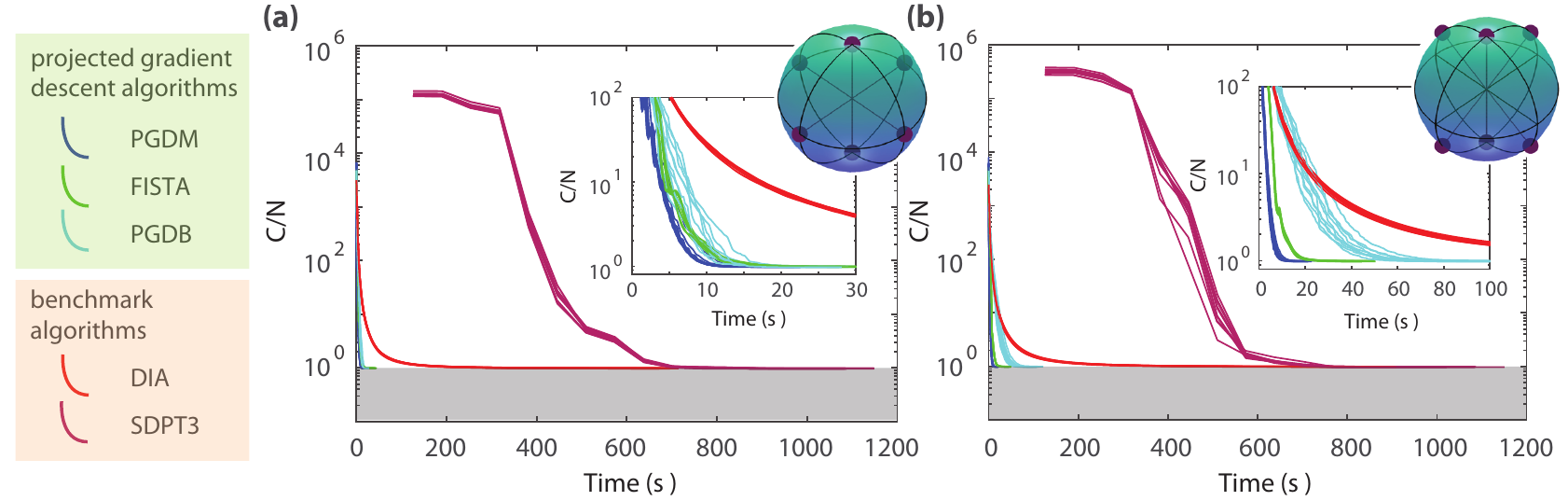} 
  \caption{\footnotesize{Typical curves of the cost function versus running time.  These simulations are performed on six-qubit systems using (a) Pauli measurements as indicated by the vectors on the Bloch sphere and (b) an ill-conditioned measurement matrix.  The global minimum of the negative log-likelihood is expected to be at $\mathcal C/N \approx 1$, around the top of the grey regions. Only for PGDM and FISTA can the cost function go up as a function of iteration number before reaching the ML state. The total running time for PGDB correlates highly with the measurement matrix condition number. }}
  \label{fig:cost}
\end{figure*}

\begin{figure*}
  \centering
  \includegraphics[]{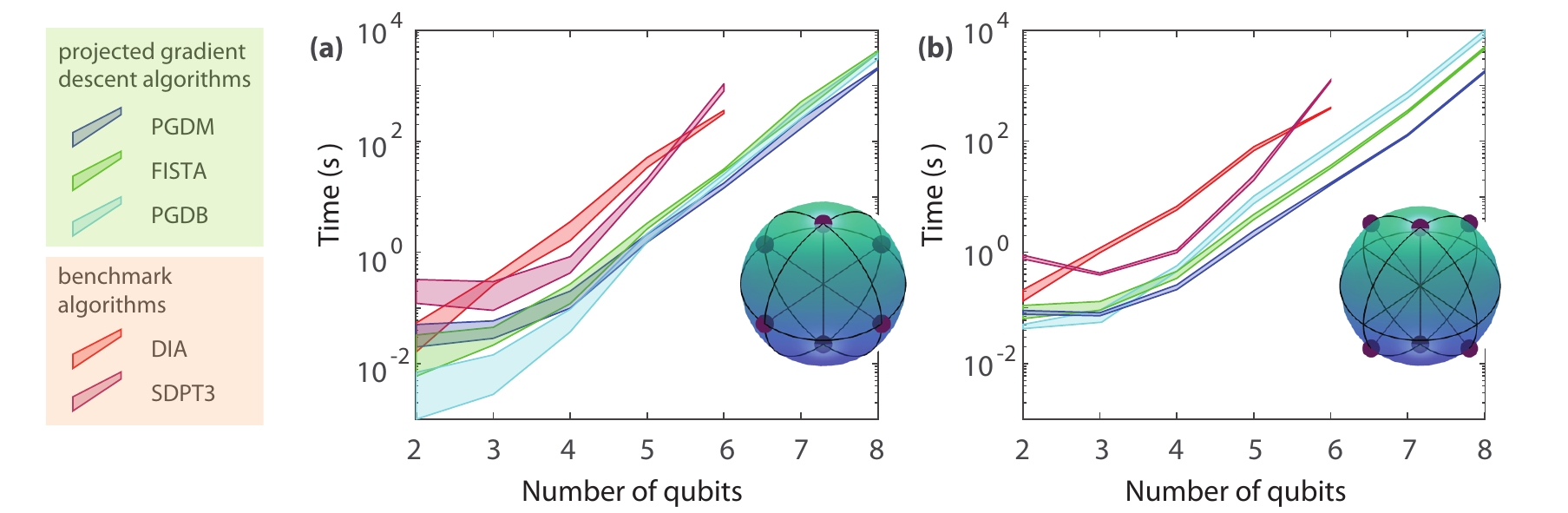} 
  \caption{\footnotesize{Average running time to reach (sufficiently close to) $\rho_\text{ML}$ for various system sizes. The measurement matrix is made of (a) Pauli measurements  and (b) bases relatively close to each other, as indicated on the inset. The colored areas are bounded by one standard deviation above and below the mean running time. The gradient-based techniques have a computational complexity of $O(N d^2)$ while SDPT3 converges in time $O(N^2 d^2)$.}}
  \label{fig:dim}
\end{figure*}

\subsubsection{Projected gradient descent with backtracking }
 This PGD method has recently been applied to quantum tomography simulations in Ref.~\cite{Goncalves:2016}. A similar variant has been studied in Ref.~\cite{Shang:2016tva}, where the authors report on a hybrid method between the DIA and PGD. The PGDB algorithm, whose main characteristic consists of trying to find the maximum step size that reduces the negative log-likelihood,  can be written as
\begin{equation}
 \rho_{k+1} = (1-\alpha) \rho_k  + \alpha \mathcal S \left[ \rho_{k} - \nabla   \mathcal C(\rho_k)\right] ~, 
 \end{equation}
  where $\alpha$ is a parameter to be loosely optimised at each step through backtracking. Each iteration of this algorithm is guaranteed to lower the negative log-likelihood unless a stationary point is reached, in which case the stopping criterion is satisfied. 

\subsection{Benchmark algorithms}

\subsubsection{Diluted Iterative Algorithm}

The diluted iterative algorithm (DIA) is based on the gradient of the  log-likelihood function. The algorithm can be simply stated with the following two iterative equations \cite{Rehacek:2001,Rehacek:2007}
 \begin{equation}
\begin{aligned}
{R}_{k} &= -H^{-1/2} [\nabla  \mathcal C] H^{-1/2} ~, \\
 \rho_k &= \frac{(I+\epsilon  {R}_{k} )\rho_{k-1}(I+\epsilon  {R}_{k} )}{\text{Tr}[(I+\epsilon  {R}_{k} )\rho_{k-1}(I+\epsilon  {R}_{k} )]} ~,
\end{aligned}
\end{equation} 
where  $H=\sum_i \Pi_i/\sum_i p_i$. The variable $\epsilon$ is optimised at every iteration, such that it minimises the log-likelihood function, and can be implemented in various ways \cite{Rehacek:2007,Goncalves:2014ts}. The matrix  $H$ reduces to the identity (up to a constant) when all the measurement operators form a POVM. The DIA leaves the density matrix estimate $\rho_k$ positive at every iteration.

  \paragraph{}  It has been observed that the DIA converges quickly for the first few iterations and converges very slowly later \cite{Goncalves:2016, Silva:2016,Shang:2016tva,Feito:2009uy}. In the Results section, we corroborate these observations. 

\subsubsection{ {Semidefinite programming}}
As we emphasised above, quantum tomography is often equivalent to minimising a convex function over a convex set. In the field of numerical optimisation, a problem is considered effectively solved if it can be cast into this form, partly because of the powerful and efficient algorithms and software packages that are available, and also because of the guarantee of global optimality for the solutions that they find. Such a software package therefore makes a natural benchmark for quantum tomography algorithms, with the the understanding that (because of their general purpose nature) they are not optimised for full tomography and unlikely to be as fast as other methods.  We use the CVX software environment; and the SDPT3 solver, which is an example of an infeasible path-following algorithm. 

\subsection{Simulation}

We perform quantum tomography simulations on multi-qubit systems with all of the techniques mentioned in the previous section. When using canonical Pauli measurements, all of the techniques are found to work well in that they all recover $\rho_\text{ML}$. Since the simulations consistently reach high likelihoods in practice, we concentrate on the total computation time. {The exit criterion for all techniques -- except for SDPT3 whose code we do not change -- is such that when the average gradients of the last 20 iterations is sufficiently small, the optimisation procedures are terminated. }

Examples of convergence curves for 6-qubit systems using Pauli measurements are shown in Fig.~\ref{fig:cost}. The details of the implementations and simulated data are provided in the Methods section.  As already remarked in Refs.~\cite{Goncalves:2016, Silva:2016,Shang:2016tva,Feito:2009uy}, the DIA displays fast convergence in the first few iterations but requires many more to finally satisfy the exit criterion. SDPT3 converges in between only 10 and 15 iterations, but each iteration has a computational complexity of $O(N^2 d^{2})$, {rendering it slow in high dimensions}. 

 We put the tomographic techniques to the test using ill-conditioned measurement matrices \cite{Miranowicz:2014vr,Bianchetti:2010tq,Feito:2009uy}, {see Fig.~\ref{fig:cost} b)}. If the measurement operators are limited to a restricted region of the Hilbert space, the condition number of the measurement matrix is greater than unity, and the error on the final density matrix estimate will necessarily increase \cite{Miranowicz:2014vr}. {Here, the measurement matrix is built using the bases}
\begin{equation}\label{illmub}
\begin{aligned}
&[0~1];[1~0];[\cos(\beta/2),\sin(\beta/2)];[\sin(\beta/2),\cos(\beta/2)]; \\
&[\cos(\beta/2),\bold{i}\sin(\beta/2)];[\sin(\beta/2),\bold{i}\cos(\beta/2)]
\end{aligned}
\end{equation}
{with $\beta=\pi/4$ for regular canonical Pauli operators and $\beta = \pi/3$ for the ill-conditioned case; see the Bloch spheres on Fig.~\ref{fig:cost} for an illustration of these vectors}.

Gon\c calves \textit{et al.}~provide a proof of the monotonicity of $\mathcal C$ for PGDB, that is to say that the cost function never increases in this algorithm. {By contrast, PGDM and FISTA are both algorithms for which the cost function may increase from one iteration to the next, but interestingly, this tends to speed up their performance relative to PGDB with regards to the ill-conditioned measurement matrices. }

{We show the running time of each algorithm as a function of Hilbert space dimensionality (number of qubits) in Fig.~\ref{fig:dim}. The speed-up of PGDM over PGDB for high-dimensions is present in both panels of Fig.~\ref{fig:dim}, but particularly pronounced for the quantum state reconstruction task based on the ill-conditioned measurement matrices, where PGDM is about ten times faster than PGDB on average for the eight and seven-qubit cases.}

\begin{figure}
  \centering
  \includegraphics{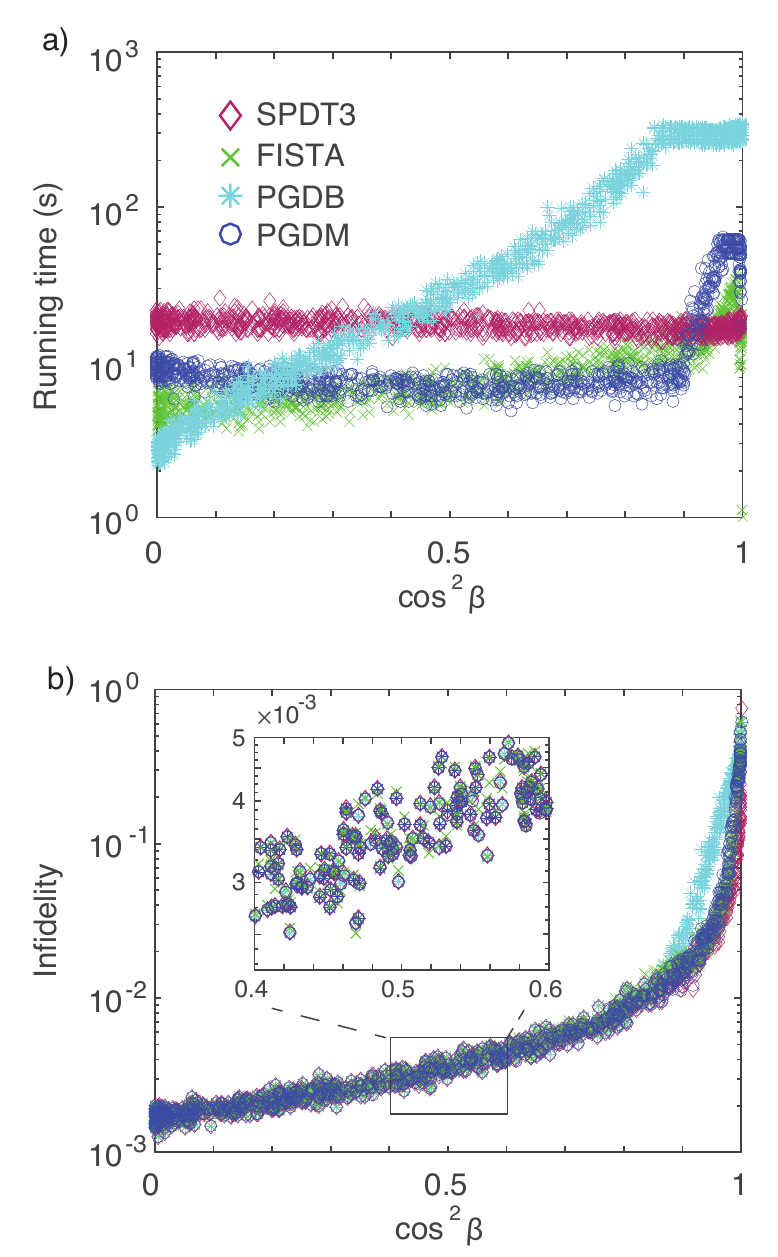} 
  \caption{\footnotesize{Scatter plot of (a) the time performance and (b) the infidelity between the recovered five-qubit states and the true states as a function of the ill-posedness of the measurement matrix. The PGDB method saturates in (a) because it reaches the maximum number of iterations set in the program. The inset in (b) shows that the methods converge towards the same fidelity, indicating that they reach $\rho_\text{ML}$. The results of FISTA differ slightly because this method oscillates around $\rho_\text{ML}$ for a large number of iterations. The number of events per outcome is set to $10^4$.}}
  \label{fig:mubness}
\end{figure}

{The running times of PGDM and FISTA algorithms are more resilient to a condition number change than PGDB, due to the fact that the number of PGDM and FISTA iterations required to reach $\rho_\text{ML}$  grows very little as a function of the measurement matrix condition number. Fig.~\ref{fig:mubness} a) illustrates this dependence.} The semidefinite programming technique does not depend on the condition number: in our simulations, SDPT3 always took about 15 iterations to reach $\rho_\text{ML}$.

There exists a monotonic relationship between ill-posedness and the accuracy of the recovered state: for a fixed number of events per measurement, the more ill-posed the measurement matrix is, the lower the fidelity between the recovered state and the true one. This relationship is illustrated in Fig.~\ref{fig:mubness}b, where the vertical axis corresponds to one minus the fidelity $f(\rho_1,\rho_2)=\text{Tr}(\sqrt{\sqrt{\rho_1}\rho_2\sqrt{\rho_1}})$ between the recovered density matrix and the true state. The extreme cases, i.e.  $\cos^2\beta=0$ and $\cos^2\beta=1$, correspond to mutually unbiased bases and a single basis measurement, respectively. We show a zoomed subset of this data around the intermediate angle, i.e. $ \cos^2 \pi/4=0.5$, that gives statistical evidence that the projected gradient descent techniques consistently converge to $\rho_\text{ML}$.

\section{Discussion}

\paragraph{} { A key advantage of the PGD techniques is their versatility. They successfully and quickly converge to the maximum likelihood state in a wide range of cases, regardless of the desired accuracy and whether the measurement matrix is well- or ill-conditioned. PGDM and FISTA -- our new PGD algorithms -- are shown to be especially well suited for ill-conditioned problems.}

\paragraph{} {Quantum tomography includes three main subfields: detector, process and state tomography. State tomography algorithms are straightforwardly transferable to process tomography, but applying the same algorithms to detector tomography with success is not trivial. The problem of detector tomography lies in characterising an unknown detector POVM from an informationally complete set of known states.  If the tomographer probes an optical detector with coherent states, the problem of detector tomography is ill-conditioned, and like the density matrix, the POVM elements must be positive-semidefinite \cite{Lundeen:2009vj}. Currently, the state-of-the-art algorithms to solve this problem are semidefinite program solvers such as SeDuMi \cite{Feito:2009uy}.  Because PGDM and FISTA perform well in the case of ill-conditioned measurement matrices, these algorithms hold great promise for optical detector tomography. One avenue for future work is the application of these two algorithms to the characterisation of detector POVMs in high dimensions.}

  \paragraph{} {In summary,  the PGD techniques have proven their worth in that they all converge towards the maximum likelihood density matrix reliably. The different PGD techniques complement each other such that PGDB is fastest in low dimensions and, according to our simulations, PGDM is fastest beyond five-qubit systems. Further, we find that the PGD techniques reach $\rho_\text{ML}$ significantly faster than the DIA and SDPT3 in the vast majority of scenarios, thus surpassing the state-of-the-art techniques with regards to assumption-free quantum state tomography. }

\subsection{Methods}

\subsubsection{Rank-1 projectors}

To reduce the memory requirements, we chose to use rank-1 measurement operators in the simulation. Instead of matrix operators, the measurements take the form of $d$-dimensional vectors. Given rank-1 projectors $\ket{\phi_i}\bra{\phi_i}$, the Born rule is written $p_i = \langle \phi_i| \rho |\phi_i\rangle$ and the measurement matrix takes the following form
\begin{equation}
\label{pureA}
\mathcal A =  \left( \begin{array}{ccc} \bra{\phi_1} \\ \vdots \\\bra{\phi_{N}}   \end{array} \right).
\end{equation}
 In this compact notation, the measurement matrix is $(N\times d)$-dimensional, thus requiring $d$ times less RAM memory compared to the full-rank case.

The gradient of the Gaussian negative log-likelihood function is compactly written as
\begin{equation}
\begin{aligned}
\nabla \log \mathcal L_G(\rho)&= 2 G^\dag \mathcal A^*,
\end{aligned} \label{grad1}
\end{equation}
where the elements of the $G$ matrix are defined:
 ${G}_{i,j} =  \mathcal  A_{i,j} (rp_i-n_i)$.  The computation of  the above gradient takes $O(N d^2)$ floating-point operations. This is the gradient that we use in the PGD algorithms for the simulations.

For all simulations in the main text, the average number of events per outcome $ r/d$ is set to $10^4$.  {Density matrices were chosen randomly in the Haar sense but always a with purity of 0.5. All simulations were performed on a single thread on an Intel Xeon Haswell processor.}

\subsubsection{PGD algorithms}

The pseudo code for PGDM, FISTA and PGDB is given in Algorithms \ref{PGDM}, \ref{FISTA} and \ref{PGDB}. The symbol $\circ$ corresponds to the Hadamard product (or element-wise multiplication).

\begin{algorithm}
\caption{PGDM}\label{PGDM}
\begin{algorithmic}[1]
\State{$k= \mathbf{0}$}
\State{Initial estimate and momentum matrix: $\rho_0=I$, $M_0=0$}
\State{$currentMagnitude = \ceil {\log_{10} \mathcal C_P(\rho_0)} $}
\State{Set step size and inertia: $\gamma=(2rd)^{-1}$, $\zeta = 0.95$}
\While{$\sum_{j=1}^{20}  |\mathcal C_P (\rho_j) - \mathcal C_P (\rho_{j-1})|> 10^{-5}$ }
\State{Projection: $\rho_k = \mathcal{S} (\rho_{k})$}
\State{Estimate probabilities: $\mathbf{p_k} = \sum_j [ \mathcal A \circ ( \mathcal  A^* {\rho_k})]_{i,j}$}
\State{Calculate log-likelihood:  $  \mathcal C_P (\rho_k)=  \boldsymbol{\nu}^T \boldsymbol{\nu}/N $}
\State{Compute gradient: $\nabla \mathcal C_G (\rho_k)= 2 G^\dag  \mathcal A^*$}
\State{$ currentMagnitude = \ceil {\log_{10} \mathcal C_P(\rho_k)} $)}
\If{$ currentMagnitude < previousMagnitude$} %ceil(log10(costFunction(it)))< orderMagnitudeCost }
\State{Update inertia: $\zeta_k = (1-(1-\zeta_k)*0.95)$}
\State{$previousMagnitude=currentMagnitude$}
\EndIf
\State{Update momentum: ${M_{k+1}} =  \zeta_k {M_{k}} - \gamma\nabla  \mathcal C (\rho_k) $}
\State{Update density matrix: $\rho_{k+1} = \rho_k + {M_{k+1}} $}
\State{$k = k+1$}
\EndWhile
\State{Final projection: $\rho_\text{final} = \mathcal{S} (\rho_{k+1})$}
\State{\textbf{Return} $ \rho_\text{final}$}
\end{algorithmic}
\end{algorithm}

%
% \rho_{k+1} = \mathcal S\left[\rho_k + \frac{k-2}{k+1}(\rho_k-\rho_{k-1})  - \delta\nabla  \mathcal C(\rho_k) \right] 

\begin{algorithm}
\caption{FISTA}\label{FISTA}
\begin{algorithmic}[1]
\State{$k= \mathbf{0}$}
\State{Initial estimate and momentum matrix: $\rho_0=I$, $M_0=0$}
\State{Set step size: $\delta = (10 d)^{-1}$ }
\While{$\sum_{j=1}^{20}  |\mathcal C_P (\rho_j) - \mathcal C_P (\rho_{j-1})|> 10^{-6}$ }
\State{Projection: $\rho_k = \mathcal{S} (\rho_{k})$}
\State{Estimate probabilities: $\mathbf{p_k} = \sum_j [ \mathcal A \circ ( \mathcal  A^* {\rho_k})]_{i,j}$}
\State{Calculate log-likelihood:  $  \mathcal C_P (\rho_k)=  \boldsymbol{\nu}^T \boldsymbol{\nu}/N $}
\State{Compute gradient: $\nabla \mathcal C_G (\rho_k)= 2 G^\dag  \mathcal A^*$}
\State{$\rho_{k+1} = \rho_k + (k-2)(\rho_k-\rho_{k-1})(k+1)^{-1}  - \delta\nabla  \mathcal C_G(\rho_k) $}
\State{$k = k+1$}
\EndWhile
\State{Final projection: $\rho_\text{final} = \mathcal{S} (\rho_{k+1})$}
\State{\textbf{Return} $ \rho_\text{final}$}
\end{algorithmic}
\end{algorithm}

\begin{algorithm}
\caption{PGDB }\label{PGDB}
\begin{algorithmic}[1]
\State{$k= \mathbf{0}$}
\State{Initial estimate and momentum matrix: $\rho_0=I$, $M_0=0$}
\State{Set metaparameters: $\mu = 1,~\ell = 10^{-4}$}
\While{$\sum_{j=1}^{20}  |\mathcal C (\rho_j) - \mathcal C (\rho_{j-1})|>10^{-5}$ }
\State{Projection: $\rho_k = \mathcal{S} (\rho_{k})$}
\State{Estimate probabilities: $\mathbf{p_k}_i = \sum_j [ \mathcal A \circ ( \mathcal  A^* {\rho_k})]_{i,j}$}
\State{Calculate log-likelihood:  $  \mathcal C (\rho_k)=  \boldsymbol{\nu}^T \boldsymbol{\nu}/N $}
\State{Compute gradient: $\nabla \mathcal C (\rho_k)= 2 P^\dag  \mathcal A^*$}
\State{$\rho_k' = \mathcal S (\rho_k - \mu^{-1} \nabla \mathcal C_G(\rho_k))$}
\State{$D = \rho_k'-\rho_k$}
\State{Line search initialisation: $\alpha_k = 1$}
\State{$\mathcal C_G'(\rho_k) = \mathcal C_G(\rho_k)+\ell \alpha_k \text{Tr}[D\nabla \mathcal C_G(\rho_k)]$}
\While{$\mathcal C_G(\rho_k+\alpha_k D) > \mathcal C_G'(\rho_k)$}
\State{Line search: $\alpha_k = \alpha_k / 2$}
\State{$\mathcal C_G'(\rho_k) = \mathcal C_G(\rho_k)+\ell \alpha_k \text{Tr}[D\nabla \mathcal C_G(\rho_k)]$}
\EndWhile
\State{Update density matrix: $\rho_{k+1} = \rho_k + \alpha_k D_k$}
\State{$k = k+1$}
\EndWhile
\State{Final projection: $\rho_\text{final} = \mathcal{S} (\rho_{k+1})$}
\State{\textbf{Return} $ \rho_\text{final}$}
\end{algorithmic}
\end{algorithm}

\subsubsection{Acknowledgements}

E.B.~acknowledges the financial support of the FQRNT, grant \#176729. J.L.~acknowledges the financial support of the Engineering and Physical Sciences Research Council (EPSRC, UK, Grants EP/M006514/1 and EP/M01326X/1).  G.C.K.~was supported by the Royal Commission for the Exhibition of 1851.   E.M.G.~acknowledges support from the Royal Society of Edinburgh and the Scottish Government.


\begin{thebibliography}{29}
\expandafter\ifx\csname natexlab\endcsname\relax\def\natexlab#1{#1}\fi
\expandafter\ifx\csname bibnamefont\endcsname\relax
  \def\bibnamefont#1{#1}\fi
\expandafter\ifx\csname bibfnamefont\endcsname\relax
  \def\bibfnamefont#1{#1}\fi
\expandafter\ifx\csname citenamefont\endcsname\relax
  \def\citenamefont#1{#1}\fi
\expandafter\ifx\csname url\endcsname\relax
  \def\url#1{\texttt{#1}}\fi
\expandafter\ifx\csname urlprefix\endcsname\relax\def\urlprefix{URL }\fi
\providecommand{\bibinfo}[2]{#2}
\providecommand{\eprint}[2][]{\url{#2}}

\bibitem[{\citenamefont{Gao et~al.}(2010)\citenamefont{Gao, Lu, Yao, Xu,
  G{\"u}hne, Goebel, Chen, Peng, Chen, and Pan}}]{Gao:2010ww}
\bibinfo{author}{\bibfnamefont{W.-B.} \bibnamefont{Gao}},
  \bibinfo{author}{\bibfnamefont{C.-Y.} \bibnamefont{Lu}},
  \bibinfo{author}{\bibfnamefont{X.-C.} \bibnamefont{Yao}},
  \bibinfo{author}{\bibfnamefont{P.}~\bibnamefont{Xu}},
  \bibinfo{author}{\bibfnamefont{O.}~\bibnamefont{G{\"u}hne}},
  \bibinfo{author}{\bibfnamefont{A.}~\bibnamefont{Goebel}},
  \bibinfo{author}{\bibfnamefont{Y.-A.} \bibnamefont{Chen}},
  \bibinfo{author}{\bibfnamefont{C.-Z.} \bibnamefont{Peng}},
  \bibinfo{author}{\bibfnamefont{Z.-B.} \bibnamefont{Chen}}, \bibnamefont{and}
  \bibinfo{author}{\bibfnamefont{J.-W.} \bibnamefont{Pan}},
  \bibinfo{journal}{Nature Phys.} \textbf{\bibinfo{volume}{6}},
  \bibinfo{pages}{331} (\bibinfo{year}{2010}).

\bibitem[{\citenamefont{Schindler et~al.}(2011)\citenamefont{Schindler,
  Barreiro, Monz, Nebendahl, Nigg, Chwalla, Hennrich, and
  Blatt}}]{Schindler:2011}
\bibinfo{author}{\bibfnamefont{P.}~\bibnamefont{Schindler}},
  \bibinfo{author}{\bibfnamefont{J.~T.} \bibnamefont{Barreiro}},
  \bibinfo{author}{\bibfnamefont{T.}~\bibnamefont{Monz}},
  \bibinfo{author}{\bibfnamefont{V.}~\bibnamefont{Nebendahl}},
  \bibinfo{author}{\bibfnamefont{D.}~\bibnamefont{Nigg}},
  \bibinfo{author}{\bibfnamefont{M.}~\bibnamefont{Chwalla}},
  \bibinfo{author}{\bibfnamefont{M.}~\bibnamefont{Hennrich}}, \bibnamefont{and}
  \bibinfo{author}{\bibfnamefont{R.}~\bibnamefont{Blatt}},
  \bibinfo{journal}{Science} \textbf{\bibinfo{volume}{332}},
  \bibinfo{pages}{1059} (\bibinfo{year}{2011}).

\bibitem[{\citenamefont{Yao et~al.}(2012{\natexlab{a}})\citenamefont{Yao, Wang,
  Xu, Lu, Pan, Bao, Peng, Lu, Chen, and Pan}}]{Yao:2012uy}
\bibinfo{author}{\bibfnamefont{X.-C.} \bibnamefont{Yao}},
  \bibinfo{author}{\bibfnamefont{T.-X.} \bibnamefont{Wang}},
  \bibinfo{author}{\bibfnamefont{P.}~\bibnamefont{Xu}},
  \bibinfo{author}{\bibfnamefont{H.}~\bibnamefont{Lu}},
  \bibinfo{author}{\bibfnamefont{G.-S.} \bibnamefont{Pan}},
  \bibinfo{author}{\bibfnamefont{X.-H.} \bibnamefont{Bao}},
  \bibinfo{author}{\bibfnamefont{C.-Z.} \bibnamefont{Peng}},
  \bibinfo{author}{\bibfnamefont{C.-Y.} \bibnamefont{Lu}},
  \bibinfo{author}{\bibfnamefont{Y.-A.} \bibnamefont{Chen}}, \bibnamefont{and}
  \bibinfo{author}{\bibfnamefont{J.-W.} \bibnamefont{Pan}},
  \bibinfo{journal}{Nature Comm.} \textbf{\bibinfo{volume}{6}},
  \bibinfo{pages}{225} (\bibinfo{year}{2012}{\natexlab{a}}).

\bibitem[{\citenamefont{Bolduc et~al.}(2016)\citenamefont{Bolduc, Gariepy, and
  Leach}}]{Bolduc:2016}
\bibinfo{author}{\bibfnamefont{E.}~\bibnamefont{Bolduc}},
  \bibinfo{author}{\bibfnamefont{G.}~\bibnamefont{Gariepy}}, \bibnamefont{and}
  \bibinfo{author}{\bibfnamefont{J.}~\bibnamefont{Leach}},
  \bibinfo{journal}{Nature Comm.} \textbf{\bibinfo{volume}{7}}
  (\bibinfo{year}{2016}).

\bibitem[{\citenamefont{Malik et~al.}(2014)\citenamefont{Malik, Mirhosseini,
  Lavery, Leach, Padgett, and Boyd}}]{Malik:2013}
\bibinfo{author}{\bibfnamefont{M.}~\bibnamefont{Malik}},
  \bibinfo{author}{\bibfnamefont{M.}~\bibnamefont{Mirhosseini}},
  \bibinfo{author}{\bibfnamefont{M.}~\bibnamefont{Lavery}},
  \bibinfo{author}{\bibfnamefont{J.}~\bibnamefont{Leach}},
  \bibinfo{author}{\bibfnamefont{M.~J.} \bibnamefont{Padgett}},
  \bibnamefont{and} \bibinfo{author}{\bibfnamefont{R.~W.} \bibnamefont{Boyd}},
  \bibinfo{journal}{Nat Commun} \textbf{\bibinfo{volume}{5}},
  \bibinfo{pages}{3115} (\bibinfo{year}{2014}).

\bibitem[{\citenamefont{Yao et~al.}(2012{\natexlab{b}})\citenamefont{Yao, Wang,
  Xu, Lu, Pan, Bao, Peng, Lu, Chen, and Pan}}]{Wong:2012}
\bibinfo{author}{\bibfnamefont{X.~C.} \bibnamefont{Yao}},
  \bibinfo{author}{\bibfnamefont{T.~X.} \bibnamefont{Wang}},
  \bibinfo{author}{\bibfnamefont{P.}~\bibnamefont{Xu}},
  \bibinfo{author}{\bibfnamefont{H.}~\bibnamefont{Lu}},
  \bibinfo{author}{\bibfnamefont{G.~S.} \bibnamefont{Pan}},
  \bibinfo{author}{\bibfnamefont{X.~H.} \bibnamefont{Bao}},
  \bibinfo{author}{\bibfnamefont{C.-Z.} \bibnamefont{Peng}},
  \bibinfo{author}{\bibfnamefont{C.-Y.} \bibnamefont{Lu}},
  \bibinfo{author}{\bibfnamefont{Y.-A.} \bibnamefont{Chen}}, \bibnamefont{and}
  \bibinfo{author}{\bibfnamefont{J.-W.} \bibnamefont{Pan}},
  \bibinfo{journal}{Nature Comm.} \textbf{\bibinfo{volume}{6}},
  \bibinfo{pages}{225} (\bibinfo{year}{2012}{\natexlab{b}}).

\bibitem[{\citenamefont{{\v{R}}eh{\'a}{\v c}ek
  et~al.}(2001)\citenamefont{{\v{R}}eh{\'a}{\v c}ek, Hradil, and Je{\v
  z}ek}}]{Rehacek:2001}
\bibinfo{author}{\bibfnamefont{J.}~\bibnamefont{{\v{R}}eh{\'a}{\v c}ek}},
  \bibinfo{author}{\bibfnamefont{Z.}~\bibnamefont{Hradil}}, \bibnamefont{and}
  \bibinfo{author}{\bibfnamefont{M.}~\bibnamefont{Je{\v z}ek}},
  \bibinfo{journal}{Phys Rev A} \textbf{\bibinfo{volume}{63}},
  \bibinfo{pages}{040303} (\bibinfo{year}{2001}).

\bibitem[{\citenamefont{{\v{R}}eh{\'a}{\v c}ek
  et~al.}(2007)\citenamefont{{\v{R}}eh{\'a}{\v c}ek, Hradil, Knill, and
  Lvovsky}}]{Rehacek:2007}
\bibinfo{author}{\bibfnamefont{J.}~\bibnamefont{{\v{R}}eh{\'a}{\v c}ek}},
  \bibinfo{author}{\bibfnamefont{Z.}~\bibnamefont{Hradil}},
  \bibinfo{author}{\bibfnamefont{E.}~\bibnamefont{Knill}}, \bibnamefont{and}
  \bibinfo{author}{\bibfnamefont{A.~I.} \bibnamefont{Lvovsky}},
  \bibinfo{journal}{Phys Rev A} \textbf{\bibinfo{volume}{75}},
  \bibinfo{pages}{042108} (\bibinfo{year}{2007}).

\bibitem[{\citenamefont{Grant et~al.}(2008)\citenamefont{Grant, Boyd, and
  Ye}}]{Grant:2008}
\bibinfo{author}{\bibfnamefont{M.}~\bibnamefont{Grant}},
  \bibinfo{author}{\bibfnamefont{S.}~\bibnamefont{Boyd}}, \bibnamefont{and}
  \bibinfo{author}{\bibfnamefont{Y.}~\bibnamefont{Ye}} (\bibinfo{year}{2008}).

\bibitem[{\citenamefont{Gon{\c c}alves et~al.}(2016)\citenamefont{Gon{\c
  c}alves, Gomes-Ruggiero, and Lavor}}]{Goncalves:2016}
\bibinfo{author}{\bibfnamefont{D.~S.} \bibnamefont{Gon{\c c}alves}},
  \bibinfo{author}{\bibfnamefont{M.~A.} \bibnamefont{Gomes-Ruggiero}},
  \bibnamefont{and} \bibinfo{author}{\bibfnamefont{C.}~\bibnamefont{Lavor}},
  \bibinfo{journal}{Optimization Methods and Software}
  \textbf{\bibinfo{volume}{31}}, \bibinfo{pages}{328} (\bibinfo{year}{2016}).

\bibitem[{\citenamefont{Silva et~al.}(2016)\citenamefont{Silva, Glancy, and
  Vasconcelos}}]{Silva:2016}
\bibinfo{author}{\bibfnamefont{G.~B.} \bibnamefont{Silva}},
  \bibinfo{author}{\bibfnamefont{S.}~\bibnamefont{Glancy}}, \bibnamefont{and}
  \bibinfo{author}{\bibfnamefont{H.~M.} \bibnamefont{Vasconcelos}},
  \bibinfo{journal}{arXiv preprint arXiv:1604.00321}  (\bibinfo{year}{2016}).

\bibitem[{\citenamefont{Toh et~al.}(1999)\citenamefont{Toh, Todd, and
  T{\"u}t{\"u}nc{\"u}}}]{Toh:1999}
\bibinfo{author}{\bibfnamefont{K.-C.} \bibnamefont{Toh}},
  \bibinfo{author}{\bibfnamefont{M.~J.} \bibnamefont{Todd}}, \bibnamefont{and}
  \bibinfo{author}{\bibfnamefont{R.~H.} \bibnamefont{T{\"u}t{\"u}nc{\"u}}},
  \bibinfo{journal}{Optimization Methods and Software}
  \textbf{\bibinfo{volume}{11}}, \bibinfo{pages}{545} (\bibinfo{year}{1999}).

\bibitem[{\citenamefont{Sturm}(1999)}]{Sturm:1999}
\bibinfo{author}{\bibfnamefont{J.~F.} \bibnamefont{Sturm}},
  \bibinfo{journal}{Optimization Methods and Software}
  \textbf{\bibinfo{volume}{11}}, \bibinfo{pages}{625} (\bibinfo{year}{1999}).

\bibitem[{\citenamefont{Smolin et~al.}(2012)\citenamefont{Smolin, Gambetta, and
  Smith}}]{Smolin:2012vp}
\bibinfo{author}{\bibfnamefont{J.~A.} \bibnamefont{Smolin}},
  \bibinfo{author}{\bibfnamefont{J.~M.} \bibnamefont{Gambetta}},
  \bibnamefont{and} \bibinfo{author}{\bibfnamefont{G.}~\bibnamefont{Smith}},
  \bibinfo{journal}{Phys Rev Lett} \textbf{\bibinfo{volume}{108}},
  \bibinfo{pages}{070502} (\bibinfo{year}{2012}).

\bibitem[{\citenamefont{Guo et~al.}(2016)\citenamefont{Guo, Zhong, Tian, Dong,
  Qi, Li, Wang, Nori, Xiang, Li et~al.}}]{Guo:2016tz}
\bibinfo{author}{\bibfnamefont{Z.~H.} \bibnamefont{Guo}},
  \bibinfo{author}{\bibfnamefont{H.-S.} \bibnamefont{Zhong}},
  \bibinfo{author}{\bibfnamefont{Y.}~\bibnamefont{Tian}},
  \bibinfo{author}{\bibfnamefont{D.}~\bibnamefont{Dong}},
  \bibinfo{author}{\bibfnamefont{B.}~\bibnamefont{Qi}},
  \bibinfo{author}{\bibfnamefont{L.}~\bibnamefont{Li}},
  \bibinfo{author}{\bibfnamefont{Y.}~\bibnamefont{Wang}},
  \bibinfo{author}{\bibfnamefont{F.}~\bibnamefont{Nori}},
  \bibinfo{author}{\bibfnamefont{G.-Y.} \bibnamefont{Xiang}},
  \bibinfo{author}{\bibfnamefont{C.-F.} \bibnamefont{Li}},
  \bibnamefont{et~al.}, \bibinfo{journal}{New J. Phys.}
  \textbf{\bibinfo{volume}{18}}, \bibinfo{pages}{083036}
  (\bibinfo{year}{2016}).

\bibitem[{\citenamefont{Shang et~al.}(2016)\citenamefont{Shang, Zhang, and
  Ng}}]{Shang:2016tva}
\bibinfo{author}{\bibfnamefont{J.}~\bibnamefont{Shang}},
  \bibinfo{author}{\bibfnamefont{Z.}~\bibnamefont{Zhang}}, \bibnamefont{and}
  \bibinfo{author}{\bibfnamefont{H.~K.} \bibnamefont{Ng}},
  \bibinfo{journal}{arXiv preprint arXiv:1609.07881}  (\bibinfo{year}{2016}).

\bibitem[{\citenamefont{Beck and Teboulle}(2009)}]{Beck:2009tk}
\bibinfo{author}{\bibfnamefont{A.}~\bibnamefont{Beck}} \bibnamefont{and}
  \bibinfo{author}{\bibfnamefont{M.}~\bibnamefont{Teboulle}},
  \bibinfo{journal}{SIAM Journal on Imaging Sciences}
  \textbf{\bibinfo{volume}{2}}, \bibinfo{pages}{183} (\bibinfo{year}{2009}).

\bibitem[{\citenamefont{{\.{Z}}yczkowski
  et~al.}(2011)\citenamefont{{\.{Z}}yczkowski, Penson, Nechita, and
  Collins}}]{Zyczkowski:2011}
\bibinfo{author}{\bibfnamefont{K.}~\bibnamefont{{\.{Z}}yczkowski}},
  \bibinfo{author}{\bibfnamefont{K.~A.} \bibnamefont{Penson}},
  \bibinfo{author}{\bibfnamefont{I.}~\bibnamefont{Nechita}}, \bibnamefont{and}
  \bibinfo{author}{\bibfnamefont{B.}~\bibnamefont{Collins}},
  \bibinfo{journal}{Journal of Mathematical Physics}
  \textbf{\bibinfo{volume}{52}}, \bibinfo{pages}{062201}
  (\bibinfo{year}{2011}).

\bibitem[{\citenamefont{Banaszek et~al.}(2000)\citenamefont{Banaszek, D'Ariano,
  Paris, and Sacchi}}]{Banaszek:1999}
\bibinfo{author}{\bibfnamefont{K.}~\bibnamefont{Banaszek}},
  \bibinfo{author}{\bibfnamefont{G.~M.} \bibnamefont{D'Ariano}},
  \bibinfo{author}{\bibfnamefont{M.~G.~A.} \bibnamefont{Paris}},
  \bibnamefont{and} \bibinfo{author}{\bibfnamefont{M.~F.}
  \bibnamefont{Sacchi}}, \bibinfo{journal}{Phys Rev A}
  \textbf{\bibinfo{volume}{61}}, \bibinfo{pages}{10304} (\bibinfo{year}{2000}).

\bibitem[{\citenamefont{Miranowicz et~al.}(2014)\citenamefont{Miranowicz,
  Bartkiewicz, Pe{\v{r}}ina~Jr, Koashi, Imoto, and Nori}}]{Miranowicz:2014vr}
\bibinfo{author}{\bibfnamefont{A.}~\bibnamefont{Miranowicz}},
  \bibinfo{author}{\bibfnamefont{K.}~\bibnamefont{Bartkiewicz}},
  \bibinfo{author}{\bibfnamefont{J.}~\bibnamefont{Pe{\v{r}}ina~Jr}},
  \bibinfo{author}{\bibfnamefont{M.}~\bibnamefont{Koashi}},
  \bibinfo{author}{\bibfnamefont{N.}~\bibnamefont{Imoto}}, \bibnamefont{and}
  \bibinfo{author}{\bibfnamefont{F.}~\bibnamefont{Nori}},
  \bibinfo{journal}{Phys Rev A} \textbf{\bibinfo{volume}{90}},
  \bibinfo{pages}{062123} (\bibinfo{year}{2014}).

\bibitem[{\citenamefont{Feito et~al.}(2009)\citenamefont{Feito, Lundeen,
  Coldenstrodt-Ronge, Eisert, Plenio, and Walmsley}}]{Feito:2009uy}
\bibinfo{author}{\bibfnamefont{A.}~\bibnamefont{Feito}},
  \bibinfo{author}{\bibfnamefont{J.~S.} \bibnamefont{Lundeen}},
  \bibinfo{author}{\bibfnamefont{H.}~\bibnamefont{Coldenstrodt-Ronge}},
  \bibinfo{author}{\bibfnamefont{J.}~\bibnamefont{Eisert}},
  \bibinfo{author}{\bibfnamefont{M.~B.} \bibnamefont{Plenio}},
  \bibnamefont{and} \bibinfo{author}{\bibfnamefont{I.~A.}
  \bibnamefont{Walmsley}}, \bibinfo{journal}{New J. Phys.}
  \textbf{\bibinfo{volume}{11}}, \bibinfo{pages}{093038}
  (\bibinfo{year}{2009}).

\bibitem[{\citenamefont{Lundeen et~al.}(2009)\citenamefont{Lundeen, Feito,
  Coldenstrodt-Ronge, Pregnell, Silberhorn, Ralph, Eisert, Plenio, and
  Walmsley}}]{Lundeen:2009vj}
\bibinfo{author}{\bibfnamefont{J.~S.} \bibnamefont{Lundeen}},
  \bibinfo{author}{\bibfnamefont{A.}~\bibnamefont{Feito}},
  \bibinfo{author}{\bibfnamefont{H.}~\bibnamefont{Coldenstrodt-Ronge}},
  \bibinfo{author}{\bibfnamefont{K.~L.} \bibnamefont{Pregnell}},
  \bibinfo{author}{\bibfnamefont{C.}~\bibnamefont{Silberhorn}},
  \bibinfo{author}{\bibfnamefont{T.~C.} \bibnamefont{Ralph}},
  \bibinfo{author}{\bibfnamefont{J.}~\bibnamefont{Eisert}},
  \bibinfo{author}{\bibfnamefont{M.~B.} \bibnamefont{Plenio}},
  \bibnamefont{and} \bibinfo{author}{\bibfnamefont{I.~A.}
  \bibnamefont{Walmsley}}, \bibinfo{journal}{Nature Phys.}
  \textbf{\bibinfo{volume}{5}}, \bibinfo{pages}{27} (\bibinfo{year}{2009}).

\bibitem[{\citenamefont{Bianchetti et~al.}(2010)\citenamefont{Bianchetti,
  Filipp, Baur, Fink, Lang, Steffen, Boissonneault, Blais, and
  Wallraff}}]{Bianchetti:2010tq}
\bibinfo{author}{\bibfnamefont{R.}~\bibnamefont{Bianchetti}},
  \bibinfo{author}{\bibfnamefont{S.}~\bibnamefont{Filipp}},
  \bibinfo{author}{\bibfnamefont{M.}~\bibnamefont{Baur}},
  \bibinfo{author}{\bibfnamefont{J.~M.} \bibnamefont{Fink}},
  \bibinfo{author}{\bibfnamefont{C.}~\bibnamefont{Lang}},
  \bibinfo{author}{\bibfnamefont{L.}~\bibnamefont{Steffen}},
  \bibinfo{author}{\bibfnamefont{M.}~\bibnamefont{Boissonneault}},
  \bibinfo{author}{\bibfnamefont{A.}~\bibnamefont{Blais}}, \bibnamefont{and}
  \bibinfo{author}{\bibfnamefont{A.}~\bibnamefont{Wallraff}},
  \bibinfo{journal}{Phys Rev Lett} \textbf{\bibinfo{volume}{105}},
  \bibinfo{pages}{223601} (\bibinfo{year}{2010}).

\bibitem[{\citenamefont{Kaznady and James}(2009)}]{Kaznady:2009wf}
\bibinfo{author}{\bibfnamefont{M.~S.} \bibnamefont{Kaznady}} \bibnamefont{and}
  \bibinfo{author}{\bibfnamefont{D.~F.} \bibnamefont{James}},
  \bibinfo{journal}{Phys Rev A} \textbf{\bibinfo{volume}{79}},
  \bibinfo{pages}{022109} (\bibinfo{year}{2009}).

\bibitem[{\citenamefont{James et~al.}(2001)\citenamefont{James, Kwiat, Munro,
  and White}}]{James:2001ut}
\bibinfo{author}{\bibfnamefont{D.~F.} \bibnamefont{James}},
  \bibinfo{author}{\bibfnamefont{P.~G.} \bibnamefont{Kwiat}},
  \bibinfo{author}{\bibfnamefont{W.~J.} \bibnamefont{Munro}}, \bibnamefont{and}
  \bibinfo{author}{\bibfnamefont{A.~G.} \bibnamefont{White}},
  \bibinfo{journal}{Phys Rev A} \textbf{\bibinfo{volume}{64}},
  \bibinfo{pages}{052312} (\bibinfo{year}{2001}).

\bibitem[{\citenamefont{Press et~al.}(1996)\citenamefont{Press, Teukolsky,
  Vetterling, and Flannery}}]{Press:1996wea}
\bibinfo{author}{\bibfnamefont{W.~H.} \bibnamefont{Press}},
  \bibinfo{author}{\bibfnamefont{S.~A.} \bibnamefont{Teukolsky}},
  \bibinfo{author}{\bibfnamefont{W.~T.} \bibnamefont{Vetterling}},
  \bibnamefont{and} \bibinfo{author}{\bibfnamefont{B.~P.}
  \bibnamefont{Flannery}}, \emph{\bibinfo{title}{{Numerical recipes in C}}},
  vol.~\bibinfo{volume}{2} (\bibinfo{publisher}{Cambridge university press
  Cambridge}, \bibinfo{year}{1996}).

\bibitem[{\citenamefont{Michelot}(1986)}]{Michelot:1986vk}
\bibinfo{author}{\bibfnamefont{C.}~\bibnamefont{Michelot}},
  \bibinfo{journal}{Journal of Optimization Theory and Applications}
  \textbf{\bibinfo{volume}{50}}, \bibinfo{pages}{195} (\bibinfo{year}{1986}).

\bibitem[{\citenamefont{Sutskever et~al.}(2013)\citenamefont{Sutskever,
  Martens, Dahl, and Hinton}}]{Sutskever:2013}
\bibinfo{author}{\bibfnamefont{I.}~\bibnamefont{Sutskever}},
  \bibinfo{author}{\bibfnamefont{J.}~\bibnamefont{Martens}},
  \bibinfo{author}{\bibfnamefont{G.~E.} \bibnamefont{Dahl}}, \bibnamefont{and}
  \bibinfo{author}{\bibfnamefont{G.~E.} \bibnamefont{Hinton}},
  \bibinfo{journal}{ICML (3)} \textbf{\bibinfo{volume}{28}},
  \bibinfo{pages}{1139} (\bibinfo{year}{2013}).

\bibitem[{\citenamefont{Gon{\c c}alves et~al.}(2014)\citenamefont{Gon{\c
  c}alves, Gomes-Ruggiero, and Lavor}}]{Goncalves:2014ts}
\bibinfo{author}{\bibfnamefont{D.~S.} \bibnamefont{Gon{\c c}alves}},
  \bibinfo{author}{\bibfnamefont{M.~A.} \bibnamefont{Gomes-Ruggiero}},
  \bibnamefont{and} \bibinfo{author}{\bibfnamefont{C.}~\bibnamefont{Lavor}},
  \bibinfo{journal}{Quantum Information {\&} Computation}
  \textbf{\bibinfo{volume}{14}}, \bibinfo{pages}{966} (\bibinfo{year}{2014}).

\end{thebibliography}
\end{document}